# Gates for Adiabatic Quantum Computing

Richard H. Warren

**Abstract.** The goal of this paper is to introduce building blocks for adiabatic quantum algorithms. Adiabatic quantum computing uses the principle of quantum annealing, which implies that a carefully controlled energy solution is optimal and corresponds to minimizing a discrete function. The input function can be influenced by rewards and penalties to favor a solution that meets restrictions that are imposed by the problem. We show how to accomplish this influence for gates in adiabatic quantum computing, particularly the controlled-NOT gate (CNOT gate) which is fundamental to all quantum gates on two or more qubits. In addition, we adapt the Toffoli gate, the Fredkin gate, and the Hadamard gate to the Ising objective function which is a foundation for discrete optimization on D-Wave System machines. The quantum work in this paper encompasses Boolean operations, some of which are used to construct gates. We think the possible advantages of the building blocks in this paper will enhance quantum algorithms on D-Wave System computers.

AMS subject classifications: Primary 81P68, Secondary 94C10

Key words: Adiabatic quantum computing, Discrete optimization problems, Ising objective function, Gates, Penalty functions

Current Address: 403 Bluebird Crossing, Glen Mills, PA 19342, USA

Email address: richard.warren@villanova.edu

# Gates for Adiabatic Quantum Computing

Richard H. Warren

**1. Introduction**

Logical gates have a major role in directing the programming flow in a quantum circuit computer. Shor [1] used gates in his theoretical algorithms for prime factorization and discrete logarithms. Logical gates have been used in algorithms that implemented Shor's work on hardware [2-5]. These demonstrations were on quantum circuit type computers or simulators.

The story is not as strong for adiabatic quantum computing. In [6-7] logical gates were shown to be transferrable in polynomial time from quantum circuit computing to adiabatic quantum computing. Initial development of the theory for gates in adiabatic quantum computing includes [8] which has a high level view that universal logical gates can be implemented as classical two-body interactions. Paper [9] continues this high level development of gates and applies them to half adders and full adders. In the current paper we amplify this development of gates with a lower level of detail about penalty functions and Hamiltonians.

Our approach starts with the Ising objective function which is minimized in adiabatic quantum computing. We describe the two types of coefficients in the Ising function [8-10] and relate them to energy fields in the hardware. Then we move through Boolean logic to penalty functions and apply them to gates in order to influence the coefficients in the Ising function.

The common thread through these steps is the CNOT gate because all quantum gates can be generated from one-qubit gates and the quantum CNOT gate [1 page 1489].

We treat the one-bit gates in adiabatic quantum computing where they fit in the discourse. The development of the CNOT gate begins with the Exclusive Or in Section 3, advances to a penalty function in Sections 4 and 5, and concludes with a Hamiltonian in Section 5. All of this is built on the foundation of Boolean logic in Section 3 and a full treatment of penalty functions in Section 4. Similar treatment is given to the gates named for Toffoli, Fredkin, and Hadamard.

Currently, stages of a problem are combined in one Hamiltonian in adiabatic quantum computing. Then all stages are optimized simultaneously. As a result, a decision in an early stage is not able to influence the computation or path in a later stage.

A significant feature in this paper is the representation of stages of a problem by iterative Hamiltonians, each executed separately in an adiabatic quantum computer and controlled overall by an executive. Iterative Hamiltonians are an ideal way to incorporate gates. This allows a gate to exercise control by being executed separately.

**2. Ising Model for Adiabatic Quantum Computing**

In an adiabatic quantum computer (AQC) the quantum bits (qubits) are loops of superconducting wire, the coupling between qubits is magnetic wiring, and the temperature is very close to 0 degrees Kelvin. References [10-13] describe a superconducting adiabatic quantum processor which currently has 512 qubits. Fabrication limits the number of pair-wise-coupled qubits, which in turn limits the number of variables for problems that are implemented on the computer. For example, if all variables are related to each other, then the problem size is restricted to at most 33 variables[1] when the problem is run on the current 512 qubit machine.

In [6-7] it is shown that the adiabatic quantum computing model is polynomially equivalent to the quantum circuit model which is the traditional method for quantum computation. This means that logical gates in the circuit model can be implemented in an AQC.

The theory for quantum annealing [10] implies that the qubits will achieve an optimal state of low energy when super cooled. The Ising objective function [14 page 1, 15 page XY, 18 page 1] for this optimal state is

$$\min\left(\sum_{i>j} s_i J_{ij} s_j + \sum_i h_i s_i\right) \tag{1}$$

where $i$ and $j$ are qubits, $s_i$ is the spin state of qubit $i$, either -1 or 1, $h_i$ is the energy bias for qubit $i$, and $J_{ij}$ is the coupling energy between qubits $i$ and $j$. An optimal state for (1) is an assignment of -1, 1 to the spin states $s_i$ and $s_j$ so that (1) is minimized.

The interface to the variables in an AQC is called a Hamiltonian, which is a square, symmetric matrix with a row and column for each spin variable $s_i$ that applies to the problem. The diagonal entries of the Hamiltonian are the values $h_i$ assigned to qubits. The off-diagonal entries are the values $J_{ij}$ assigned to the connections between qubits. The entries in the Hamiltonian are the coefficients of the terms from the sum of an objective function (1) and penalty functions which will be described in Section 4.

---

[1] Source is W. G. Macready, D-Wave Systems, Burnaby, BC, Canada.

Suppose we want to set a spin state $s_i$ so that it is a constant -1 or 1. Let $i$ be a qubit and $s_i$ its spin state. Let $h_i$ be the local field at $i$. Then $h_i s_i$ is a term in the Ising objective function (1). If we set $h_i$ sufficiently less than 0, then, by convention, $s_i$ will be 1 in the adiabatic minimization process; and if we set $h_i$ sufficiently greater than 0, then $s_i$ will be -1.

## 3. Boolean Applications

Let $i, j$ and $k$ be qubits in an AQC. We will use a transcribed model where $x_i, x_j, x_k \in$ {0,1}. This is convenient for Boolean algebra. If $s_i \in$ {-1, 1} and $x_i \in$ {0, 1}, the relationship $s_i = 2x_i - 1$ equates the spin state model and the Boolean model. So we may assume $x_i$, $x_j$ and $x_k$ are the values assigned to the qubits in the minimization process (1).

There are several unitary operations. For example, $\overline{x_k}$ which is the negation of $x_k$ where $\overline{x_k} = 0$ if $x_k = 1$ and $\overline{x_k} = 1$ if $x_k = 0$. Creation of the constant functions were described in the previous section. Other unitary operations are $x_j = x_k$ and $x_j = \overline{x_k}$.

Table 1

Truth tables for operations on two binary variables $x_i$ and $x_j$. Entries A – E are included to complete the Boolean operations on two independent variables.

| $x_i$ | $x_j$ | Conjunction (And) $x_i \wedge x_j$ | Disjunction (Inclusive Or) $x_i \vee x_j$ | Implication $x_i \to x_j$ | Exclusive Or $x_i \oplus x_j$ | Equivalence $x_i \equiv x_j$ | A | B | C | D | E |
|---|---|---|---|---|---|---|---|---|---|---|---|
| 1 | 1 | 1 | 1 | 1 | 0 | 1 | 0 | 0 | 1 | 0 | 0 |
| 1 | 0 | 0 | 1 | 0 | 1 | 0 | 0 | 1 | 1 | 0 | 1 |
| 0 | 1 | 0 | 1 | 1 | 1 | 0 | 0 | 1 | 0 | 1 | 0 |
| 0 | 0 | 0 | 0 | 1 | 0 | 1 | 1 | 1 | 1 | 0 | 0 |

Table 2

Expressions for Boolean logic in Table 1 are translated to penalty functions that can be implemented on an AQC. $a$ is an ancillary qubit so that $x_a = x_i x_j$ and is introduced to reduce a product of three variables to a quadratic term. This is necessary because the Ising function (1) is quadratic.

| Boolean Logic | Corresponding Equation(s) | Corresponding Penalty Function |
|---|---|---|
| $x_k = x_i$ | $x_k = x_i$ | $-2x_i x_k + x_i + x_k$ |
| $x_k = \bar{x}_i$ | $x_k = 1 - x_i$ | $2x_i x_k - x_i - x_k$ |
| $x_k = x_i \wedge x_j$ | $x_k = x_i x_j$ | $x_i x_j - 2(x_i + x_j)x_k + 3x_k$ |
| $x_k = x_i \vee x_j$ | $x_k = x_i + x_j - x_i x_j$ | $x_i x_j + (x_i + x_j)(1 - 2x_k) + x_k$ |
| $x_k = (x_i \rightarrow x_j)$ | $x_k = \bar{x}_i \vee x_j = 1 - x_i + x_i x_j$ | $4x_i x_j + 2x_i x_k - 6(x_i + x_j)x_a - 2x_k x_a - x_i - x_k + 9x_a$ |
| $x_k = x_i \oplus x_j$ | $x_k = (x_i \vee x_j) \wedge (\bar{x}_i \vee \bar{x}_j) = x_i + x_j - 2x_i x_j$ $= (x_i + x_j) \bmod 2 = (x_i - x_j) \bmod 2$ | $2x_i x_j - 2(x_i + x_j)x_k - 4(x_i + x_j)x_a + 4x_k x_a + x_i + x_j + x_k + 4x_a$. There is no quadratic penalty function using only variables $x_i, x_j,$ and $x_k$. |
| $x_k = (x_i \equiv x_j)$ | $x_k = \overline{(x_i \oplus x_j)} = 1 - x_i - x_j + 2x_i x_j$ | $2x_i x_j + 2(x_i + x_j)x_k - 4(x_i + x_j)x_a - 4x_k x_a - x_i - x_j - x_k + 8x_a$ |

    The penalty functions for ∧, ∨, and ⊕ are from [14 Table 4.1]. The penalty function for ≡ is the negation of the penalty function for ⊕ added to 4 times the penalty function for ∧. Similarly, the penalty function for → is a sum of two penalty functions. The first is derived from the algebraic equation for →. The other is 3 times the penalty function for ∧.

## 4. Penalty Functions

Expressions of Boolean logic, equations and inequalities cannot be entered directly in the form of (1) for adiabatic quantum computing. We will show how to convert them to penalty functions that represent them and can be entered into a Hamiltonian for (1).

Table 2 shows a conversion of expressions for Boolean logic to penalty functions. To verify that a penalty function represents its Boolean logic, we need to show that the penalty function yields the same value, say $v$, for each assignment of truth values to the expression of the Boolean logic, and that the penalty function yields at least $v + 1$ for each assignment of false values to the expression of the Boolean logic.

For example, consider the Boolean expression $z = \bar{x}$ and its penalty function $2xz - x - z$. The following truth table shows the outcomes for all possible values of $x$ and $z$.

| $x$ | $z$ | True or False | $2xz - x - z$ |
|---|---|---|---|
| 1 | 0 | $z = \bar{x}$ is true | -1 |
| 0 | 1 | $z = \bar{x}$ is true | -1 |
| 1 | 1 | $z = \bar{x}$ is false | 0 |
| 0 | 0 | $z = \bar{x}$ is false | 0 |

In this example, the penalty function $2xz - x - z$ rewards the truth of $z = \bar{x}$. The penalty function $2xz - x - z$ is added to an objective function and obtains its minimum when (1) is minimized. The reward can be increased by multiplying the penalty function by $m > 1$.

A constraint that is an equation can be changed to a penalty function by reversing the algebraic sign of all terms on one side of the equation and deleting the equality sign. Most likely, the minimum for the result is not the same as the solution for the constraint equation. Therefore, we square the result, simplify it with the property $x^2 = x$ for binary variables, and delete the constant term. It is easily seen that a binary solution occurs for the constraint equation if and only if its penalty function attains a global minimum at this solution.

For example, consider the constraint $z = x + y + 1$ where $x,y,z \in \{0, 1\}$. We form $(z - x - y - 1)^2 = 2(xy - xz - yz) + 3(x + y) - z + 1$. We delete the term 1 because it cannot be inserted in a Hamiltonian and does not affect where there is a minimum. The next table shows that the penalty function $2(xy - xz - yz) + 3(x + y) - z$ has a minimum exactly when its constraint equation is true.

| $x$ | $y$ | $z$ | True or False | $2(xy - xz - yz) + 3(x + y) - z$ |
|---|---|---|---|---|
| 1 | 1 | 1 | $z = x + y + 1$ is false | 3 |
| 1 | 1 | 0 | $z = x + y + 1$ is false | 8 |
| 1 | 0 | 1 | $z = x + y + 1$ is false | 0 |
| 1 | 0 | 0 | $z = x + y + 1$ is false | 3 |
| 0 | 1 | 1 | $z = x + y + 1$ is false | 0 |
| 0 | 1 | 0 | $z = x + y + 1$ is false | 3 |
| 0 | 0 | 1 | $z = x + y + 1$ is true | -1 |
| 0 | 0 | 0 | $z = x + y + 1$ is false | 0 |

Lastly, we show how to convert a constraint inequality to a constraint equation by inserting slack variables, as is done in the simplex algorithm for linear programming. We require the slack variables to be binary 0, 1. The number of slack variables is 1 less than the number of variables in the inequality. For example, we convert the constraint $x + y + z \leq 2$ where x, y and z are binary variables to $x + y + z + s + t = 2$ where s and t are slack variables. Then the equation is converted to a penalty function by the above technique.

See [14 Section 4.1.3] for a systematic method to construct constraint equations and constraint inequalities for a specific problem. It also includes techniques to derive penalty functions.

The largest degree of the terms in (1) is 2. If a term in a penalty function has degree greater than 2, the degree can be reduced by introducing an ancillary qubit that represents the product of two variables. There are pitfalls. The substitution of the ancillary qubit in quadratic terms may destroy the logic of the penalty function. Since there is no need to do this, this snare is easily avoided. Another difficulty is that the penalty function for ∧ in Table 2 must be considered, since its corresponding equation is a product of two variables.

The penalty functions for Boolean operations →, ⊕, and ≡ in Table 2 require an ancillary qubit. Such penalty functions need to be tested for false values of the dependent variables $x_k$ and $x_a$ occurring simultaneously, in addition to testing them separately.

## 5. Gates

Quantum circuit computers are often called gate-array quantum computers because gates have a major role directing computations. In this sense a gate is a basic quantum circuit operating on a small number of qubits. In fact, gates are the building blocks of quantum circuits for quantum circuit computers [15].

A logic gate is a device, ideal or real, that implements a Boolean operation. Quantum logic gates are reversible, i.e., they can be returned to their original setting. More precisely, a logic gate *L* is reversible if and only if *L* is a 1-1 map of its input onto its output. Thus, there is a unique inverse gate mapping the output of *L* onto its input.

Gates have not had a role in adiabatic quantum computing, probably because their logic does not fit the Ising model (1). We will show how to overcome this difficulty in the next paragraph.

Gates need a separate logical step in quantum computing. Fortunately, the work in [7] can be stretched so that the steps of an algorithm are implemented incrementally on an AQC. If an algorithm has *n* steps, then Hamiltonians $H_1, H_2 \cdots, H_n$ are formed. The output from executing $H_i$ is part of the input for Hamiltonian $H_{i+1}$. Thus, the AQC is executed *n* times. Ideally, an executive program manages the transition of the output from executing $H_i$ into $H_{i+1}$. Gates can be implemented in this approach and are able to exercise control in adiabatic quantum computing.

Gates require exclusive use of their qubits. The qubits assigned to the gate cannot be used for an objective function or another penalty function in the algorithm step containing the gate. Otherwise the effect of the gate is likely to be destroyed.

**5.1 Controlled-NOT gate**

The controlled-NOT gate (also called the CNOT gate or XOR gate) in a circuit model quantum computer negates the target bit if and only if the control bit is 1. The truth table for the CNOT gate is:

| Input | | Output | |
|---|---|---|---|
| Control | Target | Control | Target |
| 0 | 0 | 0 | 0 |
| 0 | 1 | 0 | 1 |
| 1 | 0 | 1 | 1 |
| 1 | 1 | 1 | 0 |

The CNOT gate is fundamentally important in a circuit model quantum computer, because all gates can be generated by one-bit gates and the CNOT gate [1 page 1489]. In particular, the Toffoli gate and the Fredkin gate, which are universal gates for reversible computation, can be generated by one-bit gates and the CNOT gate. The work in [6-7] implies that these results can be transferred from circuit model quantum computing to adiabatic quantum computing.

The truth table for Exclusive Or in Table 1 is identical to the truth table for the CNOT gate. Therefore, a natural way to implement a CNOT gate for adiabatic quantum computing is to use Exclusive Or. Thus, a CNOT gate in adiabatic quantum computing needs to act like the penalty function for $\oplus$ in Table 2.

The CNOT gate in the circuit model is composed of 2 qubits, a target and a control. According to Table 2, since the CNOT gate equates to $\oplus$, the CNOT gate in adiabatic quantum computing uses 4 qubits. They are target $i$, control $j$, result $k$ and ancillary $a$. We will form a Hamiltonian for a CNOT gate using these four qubits. Let $x_i, x_j, x_k$ and $x_a$ be their corresponding value in {0, 1}. From Table 2 the penalty function for the CNOT gate is

$$2x_i x_j - 2(x_i + x_j)x_k - 4(x_i + x_j)x_a + 4x_k x_a + x_i + x_j + x_k + 4x_a \qquad (2)$$

In (2), $x_a = x_i x_j$ and is used to reduce $x_i x_j x_k$ to a quadratic term which is required by (1). The coefficients of the linear terms in (2) are placed on the diagonal of the Hamiltonian and the coefficients of the quadratic terms are placed off the diagonal. Based on (2), the Hamiltonian for the CNOT gate is:

|   | i | j | k | a |
|---|---|---|---|---|
| i | 1 | 2 | -2 | -4 |
| j | 2 | 1 | -2 | -4 |
| k | -2 | -2 | 1 | 4 |
| a | -4 | -4 | 4 | 4 |

The AQC assigns values to $x_i, x_j, x_k$ and $x_a$ from {0, 1} so that (2) is minimized.

Going forward from a CNOT gate in an AQC, the control qubit and the result cubit contain all of the information. The target qubit and the ancillary qubit are available for new purposes in the algorithm step following a CNOT gate.

A CNOT gate is reversible in adiabatic quantum computing. After the gate is executed, if the control is applied to the result, then the original target is recovered.

### 5.2 Toffoli Gate

The Toffoli gate in the circuit model quantum computer is composed of 3 qubits. If the first two bits are 1, then the Toffoli gate negates the third bit. Otherwise, the third bit is not changed. The truth table for the Toffoli gate is:

| Input | | Output | |
|---|---|---|---|
| Control | Target | Control | Result |
| 0 | 0 | 0 | 0 |
| 0 | 0 | 1 | 0 | 0 | 1 |
| 0 | 1 | 0 | 0 | 1 | 0 |
| 0 | 1 | 1 | 0 | 1 | 1 |
| 1 | 0 | 0 | 1 | 0 | 0 |
| 1 | 0 | 1 | 1 | 0 | 1 |
| 1 | 1 | 0 | 1 | 1 | 1 |
| 1 | 1 | 1 | 1 | 1 | 0 |

The Toffoli gate is universal, in the sense that it can be used to build systems that will perform any Boolean computation [15 page 29]. Also the Toffoli gate is reversible. The Toffoli gate can be implemented in a quantum logic circuit by 6 CNOT gates in sequence and several 1 qubit gates [16]. In adiabatic quantum computing the Toffoli gate can be implemented by 1 CNOT gate and 6 qubits. They are 2 control bits $c1$ and $c2$, target bit $t$, result bit $r$, and 2 ancillary bits $a$ and $b$. Let $x_{c1}, x_{c2}, x_t, x_r, x_a$ and $x_b$ be their corresponding value in {0, 1}. We set $x_b = x_{c1}x_{c2}$ and observe that $x_b = 1$ if and only if $x_{c1} = 1 = x_{c2}$. Furthermore, if $x_b = 0$, then $x_r = x_t$. If $x_b = 1$, then $x_r = 1 - x_t$. Therefore, the truth table for the Toffoli gate can be written:

| Input | | Output | |
|---|---|---|---|
| Control $x_b$ | Target $x_t$ | Control $x_b$ | Result $x_r$ |
| 0 | 0 | 0 | 0 |
| 0 | 1 | 0 | 1 |
| 1 | 0 | 1 | 1 |
| 1 | 1 | 1 | 0 |

Since this is the truth table for the CNOT gate, we can use ⊕ in Table 2 to obtain a penalty function. It is

$$2x_bx_t - 2(x_b + x_t)x_r - 4(x_b + x_t)x_a + 4x_rx_a + x_b + x_t + x_r + 4x_a \qquad (3)$$

In (3), $x_a = x_bx_t$ and is used to reduce $x_bx_tx_r$ to a quadratic term which is required by (1).

Next we need a penalty function for $x_b = x_{c1}x_{c2}$. From Table 2 it is

$$x_{c1}x_{c2} - 2(x_{c1} + x_{c2})x_b + 3x_b \qquad (4)$$

We add (3) and (4) to obtain a function for the Toffoli gate. It is

$$-4x_ax_b + 4x_ax_r - 4x_ax_t - 2x_bx_{c1} - 2x_bx_{c2} - 2x_bx_r + 2x_bx_t + x_{c1}x_{c2} - 2x_rx_t + 4x_a + 4x_b + x_r + x_t \qquad (5)$$

The coefficients of (5) are the entries in a Hamiltonian for the Toffoli gate. It is

|   | $c_1$ | $c_2$ | $t$ | $r$ | $a$ | $b$ |
|---|---|---|---|---|---|---|
| $c_1$ |   | 1 |   |   |   | -2 |
| $c_2$ | 1 |   |   |   |   | -2 |
| $t$ |   |   | 1 | -2 | -4 | 2 |
| $r$ |   |   | -2 | 1 | 4 | -2 |
| $a$ |   |   | -4 | 4 | 4 | -4 |
| $b$ | -2 | -2 | 2 | -2 | -4 | 4 |

The binary inputs are $x_{c1}, x_{c2}$ and $x_t$. The AQC assigns values to $x_r, x_a$ and $x_b$ from {0, 1} so that (5) is minimized. The gate can be reversed by using the output as a new input.

In summary, we have streamlined the Toffoli gate. In the quantum circuit model it has 6 CNOT gates in sequence and several 1 qubit gates. In adiabatic quantum computing it has 1 CNOT gate that uses 6 qubits.

### 5.3 Fredkin Gate

The Fredkin gate in the circuit model computer is composed of 3 qubits. It exchanges the last two bits if the first bit is 1. The Fredkin gate is reversible and is universal in the sense that any logical operation can be constructed of Fredkin gates and 1 qubit gates [15 page 157]. The truth table for the Fredkin gate is:

| Input | | | Output | | |
|---|---|---|---|---|---|
| $c$ | $i$ | $j$ | $c$ | $m$ | $p$ |
| 0 | 0 | 0 | 0 | 0 | 0 |
| 0 | 0 | 1 | 0 | 0 | 1 |
| 0 | 1 | 0 | 0 | 1 | 0 |
| 0 | 1 | 1 | 0 | 1 | 1 |
| 1 | 0 | 0 | 1 | 0 | 0 |
| 1 | 0 | 1 | 1 | 1 | 0 |
| 1 | 1 | 0 | 1 | 0 | 1 |
| 1 | 1 | 1 | 1 | 1 | 1 |

We streamline the notation by allowing $c, i, j, m$ and $p$ to designate qubits and their value from {0, 1}. It is well known [17] that $m = (1 - c)i + cj$ and $p = ci + (1 - c)j$. Thus, the Fredkin gate can be described by binary equations that compute the outputs.

It can be shown that there is no quadratic constraint function for $m$ using only variables $c, i, j$ and $m$. Thus, a constraint function for $m$ must use ancillary variables. We compute $((1-c)i + cj - m)^2$ and simplify. The result is

$$-ci + cj + 2cim - 2cjm + i - 2im + m \tag{6}$$

The analogous result for equation $p$ is

$$ci - cj - 2cip + 2cjp + j - 2jp + p \tag{7}$$

Since the cubic terms do not fit the Ising model (1), we substitute ancillary variables $a = cm$ in the two cubic terms in (6) and $b = cp$ in the two cubic terms in (7). These products require penalty functions. From Table 2, the penalty functions are $cm - 2(c+m)a + 3a$ and $cp - 2(c+p)b + 3b$. We add (6) with $cm$ replaced by $a$, (7) with $cp$ replaced by $b$, and 2 times the penalty functions[2]. Then we simplify. The result is a function whose coefficients are the entries in a Hamiltonian for the Fredkin gate. The function is

$$-4ac + 2ai - 2aj - 4am - 4bc - 2bi + 2bj - 4bp + 2cm + 2cp - 2im - 2jp + 6a + 6b +$$
$$i + j + m + p \tag{8}$$

The Hamiltonian corresponding to (8) specifies a Fredkin gate. It is

|   | c | i | j | m | p | a | b |
|---|---|---|---|---|---|---|---|
| c |   |   |   | 2 | 2 | -4 | -4 |
| i |   | 1 |   | -2 |   | 2 | -2 |
| j |   |   | 1 |   | -2 | -2 | 2 |
| m | 2 | -2 |   | 1 |   | -4 |   |
| p | 2 |   | -2 |   | 1 |   | -4 |
| a | -4 | 2 | -2 | -4 |   | 6 |   |
| b | -4 | -2 | 2 |   | -4 |   | 6 |

---

[2] The penalty functions are added twice because $a$ and $b$ are substituted twice in (6) and (7). Adding them once can cause an incorrect answer.

When binary $c, i,$ and $j$ are given, then a D-Wave quantum computer can minimize over the Hamiltonian and provide binary answers $m$ and $p$ that can be read as output. The gate can be reversed by using the output as new input for $i$ and $j$.

We note that (8) and its Hamiltonian are not unique for a Fredkin gate. Suppose we substitute ancillary variables $d = im$ and $e = jm$ in the cubic terms in (6) and $f = ip$ and $g = jp$ in the cubic terms in (7). This increases the size of the Hamiltonian and decreases the range of entries in the Hamiltonian. We leave a research topic for future work: Determine this 9 X 9 Hamiltonian and investigate the physical properties of executing the two Hamiltonians on a D-Wave quantum computer.

**5.4 Hadamard Gate**

Let |0> be the first vector in a basis β for 2-dimensional space and let |1> be the second vector in β. Usually β = $\left\{\begin{bmatrix}1\\0\end{bmatrix}, \begin{bmatrix}0\\1\end{bmatrix}\right\}$ in vector notation. In the quantum circuit model, the Hadamard gate, *H*, maps |0> to $\frac{|0> + |1>}{\sqrt{2}}$ which is a superposition of the basis vectors, and *H* maps |1> to $\frac{|0> - |1>}{\sqrt{2}}$ which is another superposition. If α = a|0> + b|1> where $a^2 + b^2 = 1$, then *H*α = a*H*|0> + b*H*|1> = $\frac{(a+b)|0> + (a-b)|1>}{\sqrt{2}}$. Thus, *H* is a discrete Fourier transform. Figure 1 is a diagram of the Hadamard gate operating on basis vectors on the unit circle. The geometry of Figure 1 indicates that the Hadamard gate is change of orthogonal basis.

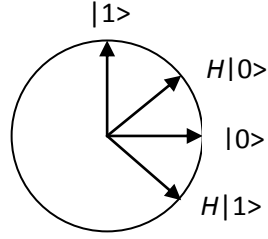

Figure 1

The Hadamard gate in matrix notation is $H = \frac{1}{\sqrt{2}}\begin{bmatrix} 1 & 1 \\ 1 & -1 \end{bmatrix}$. The operation is matrix multiplication. If $a^2 + b^2 = 1$, then $H\begin{bmatrix} a \\ b \end{bmatrix} = \frac{1}{\sqrt{2}}\begin{bmatrix} 1 & 1 \\ 1 & -1 \end{bmatrix}\begin{bmatrix} a \\ b \end{bmatrix} = \begin{bmatrix} (a+b)/\sqrt{2} \\ (a-b)/\sqrt{2} \end{bmatrix}$.

The Hadamard gate in vector notation uses ordered pairs of scalars. If $a^2 + b^2 = 1$, then $H(a,b) = \big((a+b)/\sqrt{2}, (a-b)/\sqrt{2}\big)$. In the next paragraph we will adapt this notation to the adiabatic quantum model.

In (1) the local energy field $h_i$ of qubit $i$ will have the role of a scalar for the Hadamard gate. In this sense, the Hadamard gate maps the ordered pair $(h_i, h_j)$ for qubits $i$ and $j$ to the ordered pair $\big((h_i + h_j)/\sqrt{2}, (h_i - h_j)/\sqrt{2}\big)$ where $(h_i + h_j)/\sqrt{2}$ is the local energy field for a qubit $p$ and $(h_i - h_j)/\sqrt{2}$ is the local energy field for a qubit $q$. We have assumed that $h_i^2 + h_j^2 = 1$. The following Input/Output matrix summarizes our representation of the Hadamard gate in the adiabatic quantum model.

| Input | | Output | |
|---|---|---|---|
| Qubit $i$ | Qubit $j$ | Qubit $p$ | Qubit $q$ |
| Energy $h_i$ | Energy $h_j$ | $h_p = (h_i + h_j)/\sqrt{2}$ | $h_q = (h_i - h_j)/\sqrt{2}$ |
| $h_i^2 + h_j^2 = 1$ | | | |

It is easily shown that $h_p$ and $h_q$ in the output satisfy $h_p^2 + h_q^2 = 1$. The first vector of the usual basis is represented by $(h_i, h_j) = (1,0)$, and the second vector by $(h_k, h_m) = (0,1)$. The local energy field $h_i$ depends on its qubit $i$ being in state 1, i.e., $s_i = 1$ is needed in (1). One way to accomplish this is to balance the coupling energy $J_{ik}$ between qubit $i$ and its adjacent qubits $k$. Another way is to use a penalty function $-x_i$ where $x_i \in \{0, 1\}$.

The Hadamard gate is reversible since $h_i = (h_p + h_q)/\sqrt{2}$ and $h_j = (h_p - h_q)/\sqrt{2}$.

How many qubits are needed for the Hadamard gate on a quantum adiabatic machine that uses the Ising model (1)? The Input/Output matrix shows four. If the input qubits are used for output, then the number can be reduced to two.

**Acknowledgement.** I thank the Lord Jesus who has given me ability and insight to investigate these topics, to devise the theory, and to write this paper.